# Probing phonon emission via hot carrier transport in suspended graphitic multilayers

Carlos Diaz-Pinto [a,b,†], Sungbae Lee [a,b,†], Viktor G. Hadjiev [b], Haibing Peng [a,b,*]

[a] Department of Physics and [b] the Texas Center for Superconductivity, University of Houston, Houston, Texas 77204

[†] Authors contributed equally to this work. Present address of S. Lee: General Studies Division, Gwangju Institute of Science and Technology, Gwangju 500-712, Republic of Korea.

[*] Corresponding author: haibingpeng@uh.edu

ABSTRACT

We study hot carrier transport under magnetic fields up to 15 T in suspended graphitic multilayers through differential conductance ($dI/dV$) spectroscopy. Distinct high-energy $dI/dV$ anomalies have been observed and shown to be related to intrinsic phonon-emission processes in graphite. The evolution of such $dI/dV$ anomalies under magnetic fields is further understood as a consequence of inter-Landau level cyclotron-phonon resonance scattering. The observed magneto-phonon effects not only shed light on the physical mechanisms responsible for high-current transport in graphitic systems, but also offer new perspectives for optimizing performance in graphitic nano-electronic devices.

**1.1 Introduction**

Single and multilayer graphene systems are promising for nanoelectronic applications because of their high carrier mobility and large current-carrying capacities. While electron transport studies have led to a better understanding of the physical mechanism responsible for the carrier mobility and electron scattering at low electric fields, more experiments are critically needed for elucidating the physics of transport at high electric fields. [1-7]  In particular, current saturation with increasing drain-source bias has been observed in single-layer graphene field effect transistors, [1] and it was attributed to scattering from surface phonons of the underlying $SiO_2$ substrates (distinct to carbon nanotubes [8] for which the scattering by intrinsic optical phonons is significant in causing the current saturation).  To address the role of intrinsic phonon-scattering in graphitic systems, existing experiments have mainly been directed to high drain-source bias electron transport at zero magnetic field. [1, 2, 5, 7]  However, little experiment has been carried out to study the cyclotron-phonon scattering in graphitic systems through hot carrier transport under high magnetic fields, despite the fact that it has been widely recognized as an effective probe for the magneto-phonon effect in semiconductor-based two-dimensional electron systems. [9-11]

In this work, we experimentally study the hot carrier transport under magnetic fields up to 15 T in suspended graphitic multilayers purposely designed to minimize the influence of substrate phonons.  Differential conductance ($dI/dV$) spectroscopy is performed as a function of drain-source bias $V_d$, temperature and magnetic field. Furthermore, phonon density of states and Landau level spectra for graphite are obtained theoretically to explain the experimental observation.  Our study provides direct evidence of intrinsic phonon-emission as a source of carrier scattering for high-current transport in graphitic system.

**1.2 Material and method**

In experiments, we construct two-terminal devices containing suspended graphitic multilayers (Fig. 1a) by a recently developed technique. [12]  The key point for device fabrication is to prepare liquid solutions containing a sufficient amount of thin graphitic multilayers by mechanically grinding highly-oriented pyrolytic graphite. [12]  With that, we spin-coat graphitic multilayers from the liquid solutions onto metal electrodes pre-patterned on the top thermal oxide layer (200 nm thick) of a degenerately $n$-doped Si wafer.  The gap between the source and drain electrodes is typically ~ 400 nm wide, and the electrodes (Pd/Cr) are ~50 nm thick.  By chance, a graphitic multilayer is suspended between the two elevated source and drain electrodes.  The relatively narrow gap between elevated electrodes enables us to obtain such devices (Fig. 1a) with suspended graphitic multilayers electrically connecting the source and drain.  Subsequently, the devices are annealed at a temperature of 300 ºC in Ar gas.  Electrical characterization is then carried out in a two-terminal configuration by using the degenerately doped silicon as a back gate.  The differential conductance $dI/dV$ is measured by standard lock-in technique with a small excitation voltage (100 µV or less) at a frequency 503 Hz, superimposed to a DC bias $V_d$. Magnetic fields are applied perpendicular to the graphitic multilayers with a superconducting magnet inside a cryostat.

**1.3 Results and discussion**

The inset of Fig. 1b displays a SEM image of a device with a suspended graphitic multilayer bridging electrodes. The gap between source and drain is ~ 400 nm and the width of the graphitic multilayer is ~ 900 nm. The multilayer is 18 nm thick as measured by atomic force microscopy. This device shows ~ 4 % tuning in differential conductance within a range of 50 V in applied gate voltage $V_g$ (Fig. 1b) at drain-source bias $V_d = 0$. The device is hole doped as shown also in the DC drain-source current $I_d$ as a function of $V_g$. (Fig. 1c) The Raman spectrum of the multilayer (Fig. 1d) exhibits an asymmetric G´ band related to the Bernal stacking, in contrast to the symmetric single Lorentzian G´ band observed for turbostratic multilayer graphene. [13] We estimate the stacking order coherence length $l_c$ of the graphitic multilayer using the methodology suggested in Ref. [14]. From the Raman intensity of the decomposed G'$_2$ and G'$_3$ bands (Fig. 1d), $I_{G'2}$ and $I_{G'3}$, we find $l_c = 10 + 10/(1.05-A) \approx 27$ nm, where $A = I_{G'3}/(I_{G'2} + I_{G'3})$. Hence, the estimated value of $l_c$ is comparable to the thickness of the graphitic multilayer and demonstrates single-crystal quality along the c-axis. The Raman spectrum also shows a weak D peak (Fig. 1d inset), which could be related to the submicron size of the sample (smaller than the spot size of the laser beam, ~1um in radius).

Fig. 2a presents $dI/dV$ versus $V_d$ at $V_g = 0$ for the graphitic multilayer. For devices with low contact barriers we usually observe symmetric $dI/dV$ as a function of $V_d$, while for devices with significant contact barriers (typically with conductance less than 100 μS) we often find asymmetric $dI/dV$ behaviors and upon annealing at high electrical current such devices change to symmetric $dI/dV$ behaviors after the contacts are improved. [12] The device here belongs to the latter type of devices with asymmetric contact barriers and shows a notable asymmetry in $dI/dV$ versus $V_d$. In such devices with asymmetric contact barriers, the voltage drop is mainly distributed on the contact with high barrier so that we can approximate the energy of the injecting carriers as $|eV_d|$.

At a temperature $T = 4.2$ K, two notable features are observed. First, a dip of $dI/dV$ clearly shows up at $V_d = 0$, which can be attributed to the hot electron effect. [12] In brief, electrons are almost out of thermal contact with phonons at low $V_d$ due to a weak acoustic-phonon coupling at $T = 4.2$ K, and thus they are heated up by the electric field until the energy gain is balanced by the energy loss due to electron-lattice interaction. This leads to an effective electron temperature higher than the lattice temperature, and a rise of conductance occurs near $V_d = 0$ following its temperature dependence. As $T$ increases to 50 K and above, the $dI/dV$ dip at $V_d = 0$ disappears because the electron temperature is no longer different from the lattice temperature due to enhanced electron-phonon scattering.

Second, anomalies in $dI/dV$ are observed at higher energies (> 8 meV) outside the previously discussed hot electron dip near $V_d = 0$. For the $dI/dV$ vs. $V_d$ data at $T = 4.2$ K (Fig. 2a), distinctive valleys, peaks, and changes of slope appear on the negative and positive $V_d$ side. At higher temperatures, these anomalies disappear. In addition, such $dI/dV$ anomalies are at the same positions as the gate voltage $V_g$ is varied (Fig. 2b), indicating that they are not induced by anomalies in the electronic density of states (otherwise, the positions of $dI/dV$ anomalies should shift as the chemical potential is tuned by $V_g$). We also note that such $dI/dV$ anomalies are not induced by defects scattering, since we do not observe such $dI/dV$ anomalies in samples where we purposely introduced defects through ion-beam implantation. [15] The most likely origin of the high-energy $dI/dV$ anomalies is the inelastic scattering of injected high-energy electrons through emission of phonons. [16-18] In this scenario, the probability of phonon emission processes is determined by the phonon density of states and the electron-phonon coupling strength. Via the phonon emission, an injected high-energy electron is scattered into a final electronic state (most likely, an initially unoccupied state slightly above the Fermi level). At low temperature (T = 4.2 K), the occupation probability for those relevant final electronic states with energy slightly above the Fermi level is small and thus inelastic scattering of electrons is considerable through phonon emission processes, leading to *dI/dV* peaks (valleys) at energies with minimum (maximum) phonon density of states. But as the temperature is increased to 100 K (thermal energy $k_BT \sim 9$ meV), the occupation probability is no longer small for those relevant final electronic states above the Fermi level, leading to a suppression of phonon emission processes and thus the disappearance of *dI/dV* anomalies (Fig. 2a).

Such *dI/dV* anomalies can be better visualized by subtracting the data at T= 4.2K by the data at T = 100 K (Fig. 2c). To confirm the phonon-emission picture discussed above, we also plot in Fig. 2c the phonon density of states (ph-DOS) for graphite calculated using the Density Functional Perturbation Theory. [19] A clear correlation between the ph-DOS and the processed *dI/dV* spectrum is observed: most valleys (peaks) of *dI/dV* occur at positions with maximum (minimum) ph-DOS. Notably, the minimum ph-DOS at 65 meV leads to the *dI/dV* peaks at Vd = ± 65 mV (marked by arrows) for the electron and hole, respectively. The cluster of ph-DOS peaks between 65 meV to 110 meV lead to a broad concave shape in *dI/dV* in this energy range for both the electron and hole sides, and a careful examination of *dI/dV* "fluctuations" within the broad concave features points to their correlation to the ph-DOS fluctuations (though the *dI/dV* details also depend on the electron-phonon coupling strength at different phonon energy).  Near the ph-DOS peak at 58 meV, the *dI/dV* displays a sharp drop in the electron side, indicating a strong electron-phonon coupling near this energy.  For energy lower than 50 meV, one weak peak at 47 meV and one strong peak at 20 meV in ph-DOS can be assigned to explain the observed *dI/dV* valleys in the electron side at $V_d = -45$ mV and $-18$ mV, respectively.  However, in the hole side, the *dI/dV* demonstrates anomalies as changes in *dI/dV* slopes at these two energy positions.  For a weak ph-DOS peak at 14 meV, a *dI/dV* valley is clearly observed at $V_d = +14$ mV for holes, but such a feature is not obvious for electrons, indicating different coupling strength to phonons for electrons and holes.  Overall, nearly all of the observed *dI/dV* anomalies can be explained by one-phonon emission as shown by a direct comparison with the ph-DOS, except for the *dI/dV* valleys at Vd = ± 29 mV where no hint of peak is shown in the ph-DOS.  However, this anomaly at 29 meV can be induced by the two-phonon emission process, considering the ph-DOS peak at ~14 meV.  We note that the correlation between the ph-DOS and the dI/dV spectra has been observed in multiple devices, though the dI/dV backgrounds are different for different devices and the relative strength of anomalies is different (see Supplementary Materials for the comparison between the dI/dV data and the ph-DOS for two other devices).

      Under magnetic fields, the *dI/dV* features change significantly (Fig. 3a).  First, the hot electron dip near $V_d = 0$ evolves at high magnetic fields and it is attenuated at $B = 15$ T.  The B-field induced attenuation of the *dI/dV* dip can be attributed to the intra-Landau level cyclotron-phonon scattering via low energy acoustic phonons (Fig. 3b, left panel). [9]  At higher magnetic fields, electron-acoustic phonon scattering is enhanced due to the

increased electronic density of states since the degeneracy of Landau levels is proportional to $B$. As a result, it is easier for electrons to lose their energy gain from electric fields to the lattice, and the achievable effective electron temperature is no longer as high as that at $B = 0$. Therefore, the hot electron effect at finite bias $V_d$ is diminished, leading to the attenuation of the $dI/dV$ dip at $V_d = 0$. [20]

On the other hand, the evolution of the high-energy $dI/dV$ anomalies is more complicate as the $B$ field increases. Below we mainly discuss the anomalies for the electron side (negative $V_d$) since they are more prominent than those for the hole side (positive $V_d$) (though hints for such anomalies can still be seen at similar energies in the hole side). It is worth noting that the anomaly at $V_d \sim -45$ mV persists in all magnetic fields: a slope change in $dI/dV$ is obvious at this position. Also, some features notably emerge at $V_d$ between 0 to -45 mV for some magnetic fields, which are almost aligned with the anomalies observed at $B = 0$ (marked by vertical lines in Fig. 3a). To explain the $dI/dV$ anomalies observed under magnetic fields, one may invoke the direct Landau level spectroscopy. [21-23] However, the data are inconsistent with the Landau level spectrum from two considerations: (1) the $dI/dV$ anomalies do not shift systematically to higher energies as $B$ increases; (2) No peak of $dI/dV$ can be assigned to a $B$-field-independent zeroth Landau level for graphitic systems, which has been shown to be significant [21] in scanning tunneling microscopy experiments. Further evidence to exclude the assignment of the observed $dI/dV$ anomalies to Landau level spectra comes from the differential resistance $R$ as a function of $B$ field at different drain-source bias $V_d$ (Fig. 4). For $V_d = 0$, the magneto-resistance (MR) demonstrates Shubnikov-de Haas (SdH) oscillations on top of a smooth background. But for $V_d = -10$ mV and -20 mV, the SdH oscillations are suppressed as the effective temperature is increased due to the hot electron effect, and only a positive linear MR background is observed.[24] This implies that the Landau level spectrum should be thermally smeared out for $V_d \sim$ -20 mV or higher in electron transport experiments. However, most *dI/dV* anomalies (Fig. 3) are observed at higher $V_d$ values (or higher effective temperatures) where Landau levels are thermally washed out. Therefore, we conclude that the observed *dI/dV* anomalies are not due to a direct probing of the Landau level spectrum.

Following the discussion on the phonon-emission at zero field, we can attribute the physical origin of such high-energy *dI/dV* anomalies to the inter-Landau level cyclotron-phonon scattering (Fig. 3b, right panel). Under high magnetic fields, injected energetic electrons make inter-Landau level transitions by emitting high-energy phonons, [9-11] and resonance occurs when the phonon energy matches the energy difference between the Landau levels for the initial and final electron states. Hence, a *dI/dV* anomaly should appear at specific $V_d$ values when energetic electrons are injected into a Landau level that is resonantly coupled to a lower Landau level through phonon emission. In graphitic systems, there is a unique zeroth Landau level independent of *B* field and Landau levels are unevenly spaced, [25-27] which explains the seemly "irregular" positions of the high-energy anomalies observed at different magnetic fields. For the graphitic multilayers here, the Landau level spectrum can be considered the same as graphite, except that the wave number along the c-axis ($k_z$) is quantized due to the boundary confinement. To quantitatively check the possibility of cyclotron-phonon resonance scattering, we have included in Fig. 5 the Landau level spectra under selected *B* fields relevant for the experiments, as calculated by the McClure-Inoue methods. [25-28] Considering the scattering between higher electron Landau levels and the zeroth Landau level via the emission of three-dimensional phonons, we find resonance conditions for all the observed *dI/dV* anomalies (marked in Fig. 5). In particular, we note that if a *dI/dV* anomaly is observed at a finite *B* field, it appears at *B* = 0 (as indicated by the vertical lines in Fig. 3), which further supports the physical picture of cyclotron-phonon resonance scattering (since the phonon density of states likely varies little as a function of *B* field) and rules out an explanation with the direct Landau level spectroscopy.

**1.4 Conclusions**

In summary, through hot carrier transport experiments in suspended graphitic multilayers, we have observed distinct high-energy *dI/dV* anomalies which are attributed to intrinsic electron-phonon scattering through phonon-emission process in graphitic systems. The evidence of such phonon-emission processes has been demonstrated by a direct comparison of the *dI/dV* spectrum with calculated phonon density of states for graphite. Furthermore, the evolution of high-energy *dI/dV* anomalies has been studied under magnetic fields up to 15 T, and the results can be understood as a consequence of the inter-Landau level cyclotron-phonon resonance scattering. Our demonstration of intrinsic magneto-phonon effects through hot carrier transport not only sheds light on the physical mechanisms responsible for high-current transport in graphitic systems, but also offers new perspectives for optimizing device performance for graphitic nano-electronic devices.

FIGURE CAPTIONS

**Figure 1** (a) Cross section (upper panel) and top view (lower panel) of the device scheme. (b) Tuning of the differential conductance ($dI/dV$) through the gate voltage $V_g$ with the drain-source bias $V_d = 0$ at temperature $T = 4.2$ K for a device with a graphitic multilayer bridging electrodes. Inset: SEM image of the suspended graphitic multilayer under study here (Scale bar: 1 µm). (c) DC drain-source current $I_d$ as a function of $V_g$ at fixed drain-source bias $V_d = 100$ mV at temperature $T = 4.2$ K. (d) Raman spectra of the same graphitic multilayer excited with the 514 nm Ar$^+$ laser line. Main panel: measured G´ band (symbol) and a fitting (line) to the data by a superposition of three Lorentzians, G'$_1$, G'$_2$, and G'$_3$ centered at 2687 cm$^{-1}$, 2707 cm$^{-1}$, and 2730 cm$^{-1}$, respectively. Inset: measured D and G bands.

**Figure 2** (a) Differential conductance $dI/dV$ as a function of drain-source bias $V_d$ with gate voltage $V_g = 0$ at different temperatures. (b) $dI/dV$ as a function of $V_d$ at temperature $T = 4.2$ K, measured at various gate voltage $V_g$ (from bottom to top: $V_g = -50, -30, -10, 0, +10, +30,$ and $+50$ V, respectively). For clarity, curves for different $V_g$ are stacked by a vertical offset of 8 µS in $dI/dV$ from bottom to top. (c) Bottom: the $dI/dV$ data of (a) at T = 4.2K subtracted by the data at T = 100 K. Top: theoretical values of phonon density of states (ph-DOS) for graphite. Vertical lines are eye guides to the positions of ph-DOS peaks below 50 meV.

**Figure 3** (a) Differential conductance $dI/dV$ as a function of drain-source bias $V_d$, with gate voltage $V_g = 0$ at temperature $T = 4.2$ K, measured at different magnetic fields $B$.

(All data are shown in actual values without any offset.) Vertical lines (dash) at positions $V_d = \pm 45, \pm 29$ and $\pm 18$ mV are guides to show the alignment of the $dI/dV$ anomalies at $B = 0$ with those at finite $B$ fields. (b) Schematic diagrams for intra-Landau level (left) and inter-Landau level (right) cyclotron-phonon scattering.

**Figure 4** Differential resistance $R$ (the inverse of measured $dI/dV$) as a function of magnetic fields $B$ applied perpendicular to the graphitic multilayer for selected $V_d$ from 0 to -100 mV (in a step of -10 mV shown from top to bottom) at temperature $T = 4.2$ K for the same device in Fig. 3. (For clarity, curves are stacked by a vertical offset of -0.5 K$\Omega$ from top to bottom.) SdH oscillations are observed on top of a linear MR background at $V_d = 0$, but smeared out at $V_d = -10$ and $-20$ mV. For higher $V_d$ values, the linear MR behavior persists except for the $B$ fields where high-energy anomalies appear as shown in Fig. 3. For $V_d$ from $-70$ to $-100$ mV, the data are nearly identical and the four curves overlap with each other if plotted without offset.

**Figure 5** Landau levels for $B = 3, 6, 9, 12$, and 15 T for bulk graphite, as calculated by the McClure-Inoue methods [Ref. 25-27]. The Landau levels are obtained by numerically solving Eq. (10) in Ref. 27, with relevant band parameters adopted from Ref. 28. The wave number along the c-axis, $k_z$, is shown in a unit of $2\pi/c_0$, with $c_0$ the lattice constant along c-axis. The Fermi levels ($E_F$) and energy positions for observed $dI/dV$ anomalies in Fig. 3 are indicated by horizontal lines for all $B$ fields. Landau levels are labeled according to Ref. 26. For clarity, we only display the first few Landau levels for the $\varepsilon_n^+(2)$ and $\varepsilon_n^-(1)$ Landau level series, corresponding to the electron and hole $E_3$ bands

at $B = 0$ according to Ref. 26. The Landau levels $\varepsilon_0^+(2)$ and $\varepsilon_0^-(1)$ are referred to as the zeroth electron and zeroth hole Landau levels, respectively. For each B field, a possible transition between Landau levels coupled by a phonon is marked by an arrow.

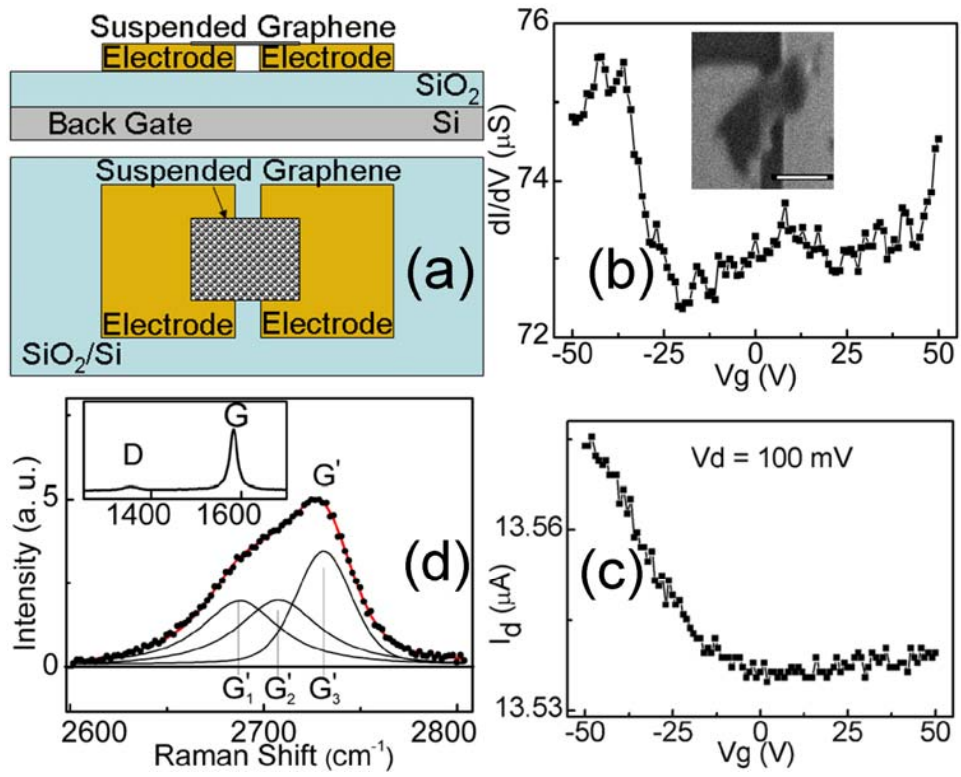

Fig. 1

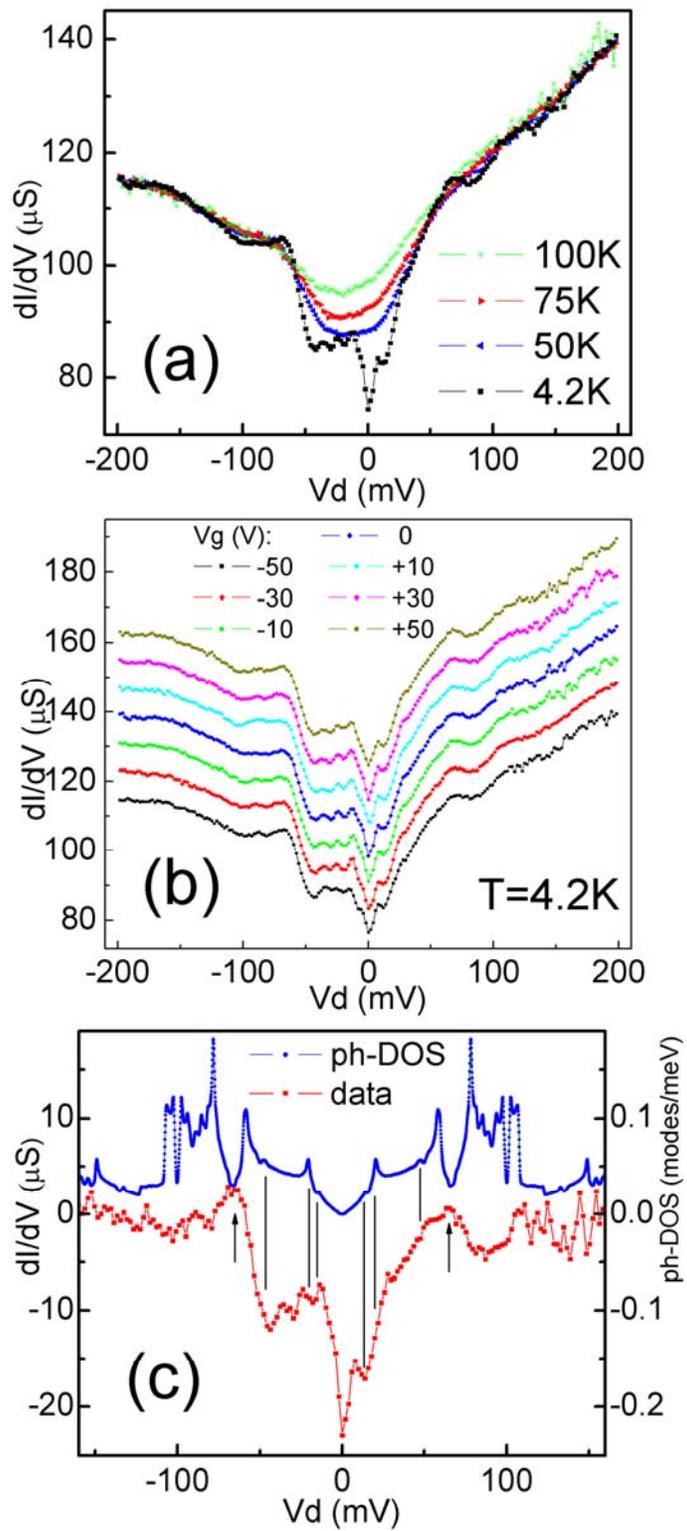

Fig. 2

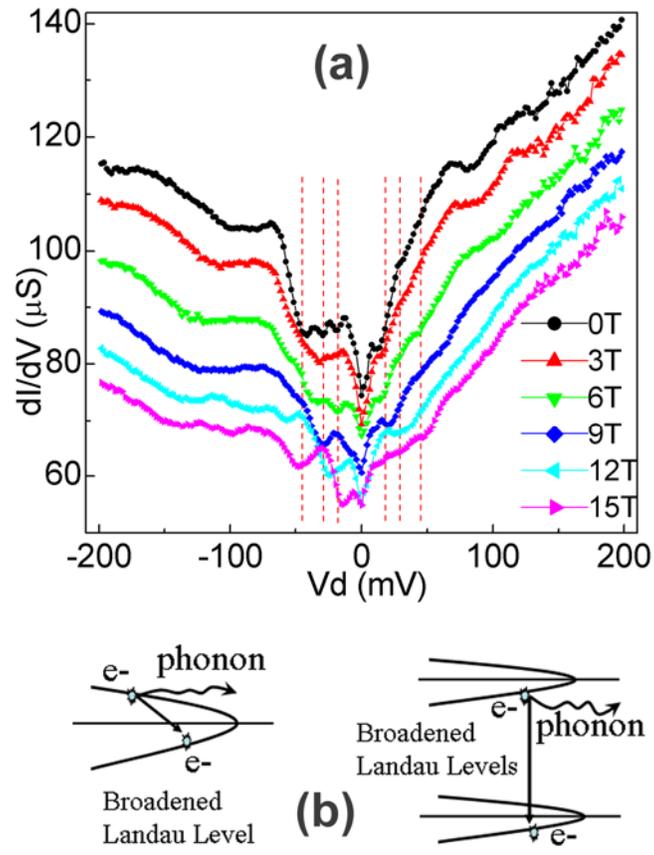

Fig. 3

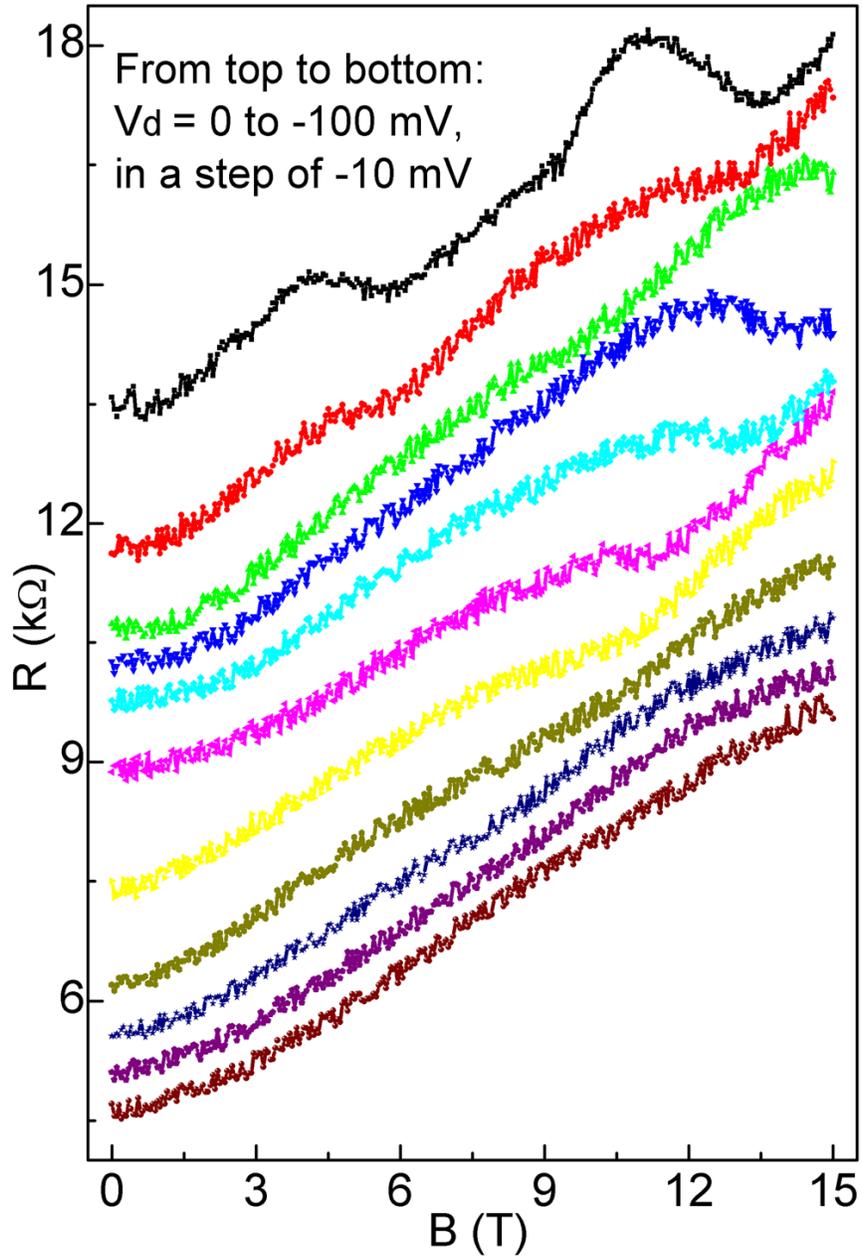

Fig. 4

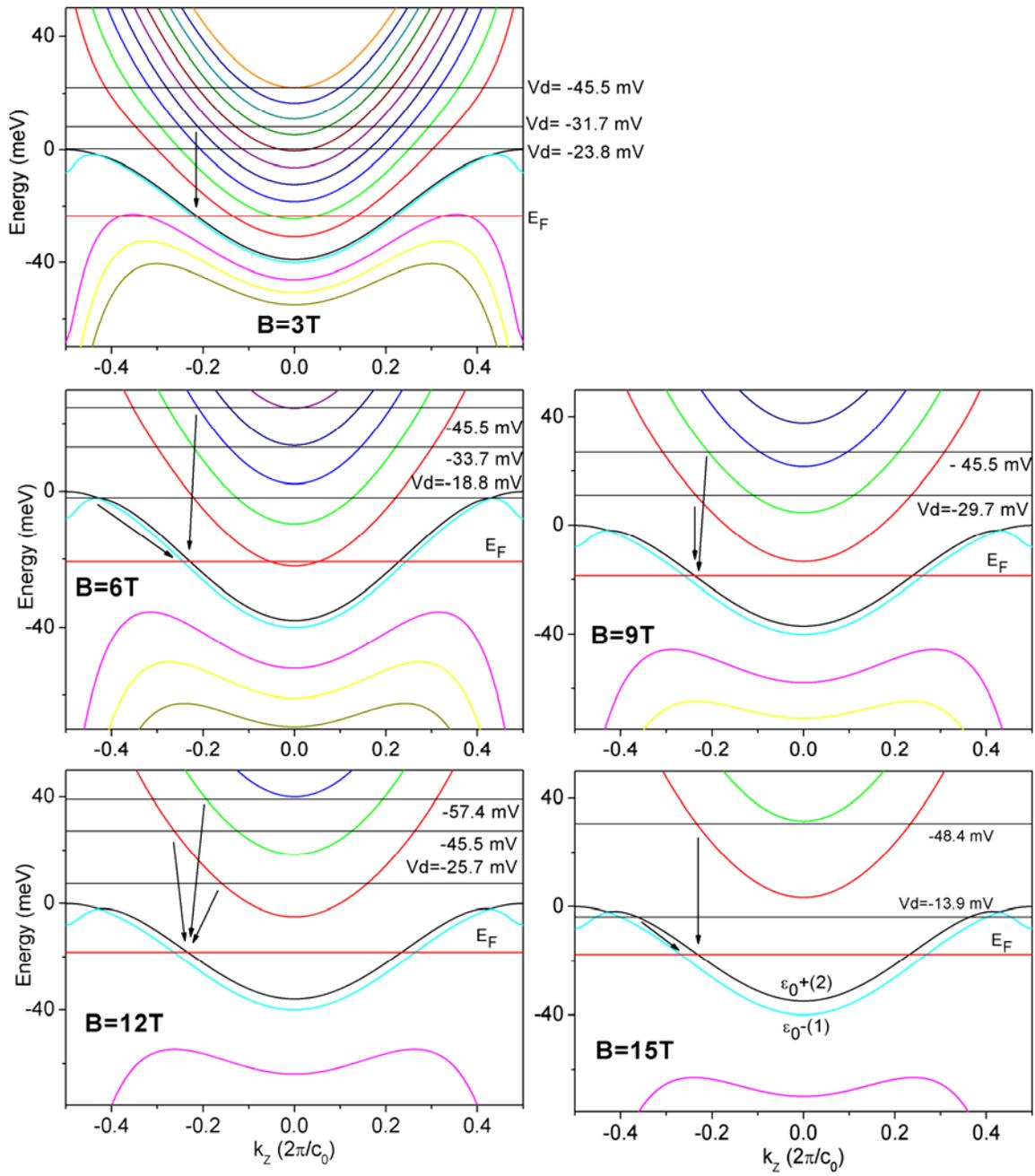

Fig. 5

# Supplementary materials for:

# Probing phonon emission via hot carrier transport in suspended graphitic multilayers

Carlos Diaz-Pinto [a,b,†], Sungbae Lee [a,b,†], Viktor G. Hadjiev [b], Haibing Peng [a,b,*]

[a] Department of Physics and [b] the Texas Center for Superconductivity, University of Houston, Houston, Texas 77204

[†] Authors contributed equally to this work. Present address of S. Lee: General Studies Division, Gwangju Institute of Science and Technology, Gwangju 500-712, Republic of Korea.

[*] Corresponding author: haibingpeng@uh.edu

## A. Calculation of the phonon dispersion and phonon density of states (ph-DOS) for graphite

**Description:** Figure 1S shows the phonon dispersion and phonon density of states (ph-DOS) of graphite calculated using the Density Functional Perturbation Theory (DFTP) [1] as implemented in the CASTEP code.[2] The first-principles energy calculations were done within the generalized-gradient approximation (GGA) with Perdew-Burke-Ernzerhof (PBE) exchange-correlation functional [3] using the norm-conserved pseudopotential plane-wave method. The non-bonding interaction between graphene planes were accounted for by dispersion corrections of the form $C_6 R^{-6}$ included in the DFT formalism by using Tkathcenko and Scheffer scheme.[4] We used a graphite crystal structure obtained from the experimental one by energy minimization and convergence tolerances $1.8 \times 10^{-7}$ eV/atom for energy, $2.0 \times 10^{-7}$ eV/Å for forces, 0.009 GPa for stresses, and $4.7 \times 10^{-10}$ Å for displacements. The phonon dispersion and ph-DOS were calculated over $23 \times 23 \times 8$ Monkhorst-Pack grid in the $k$-space along the lines connecting the high symmetry points in the Brillouin zone.

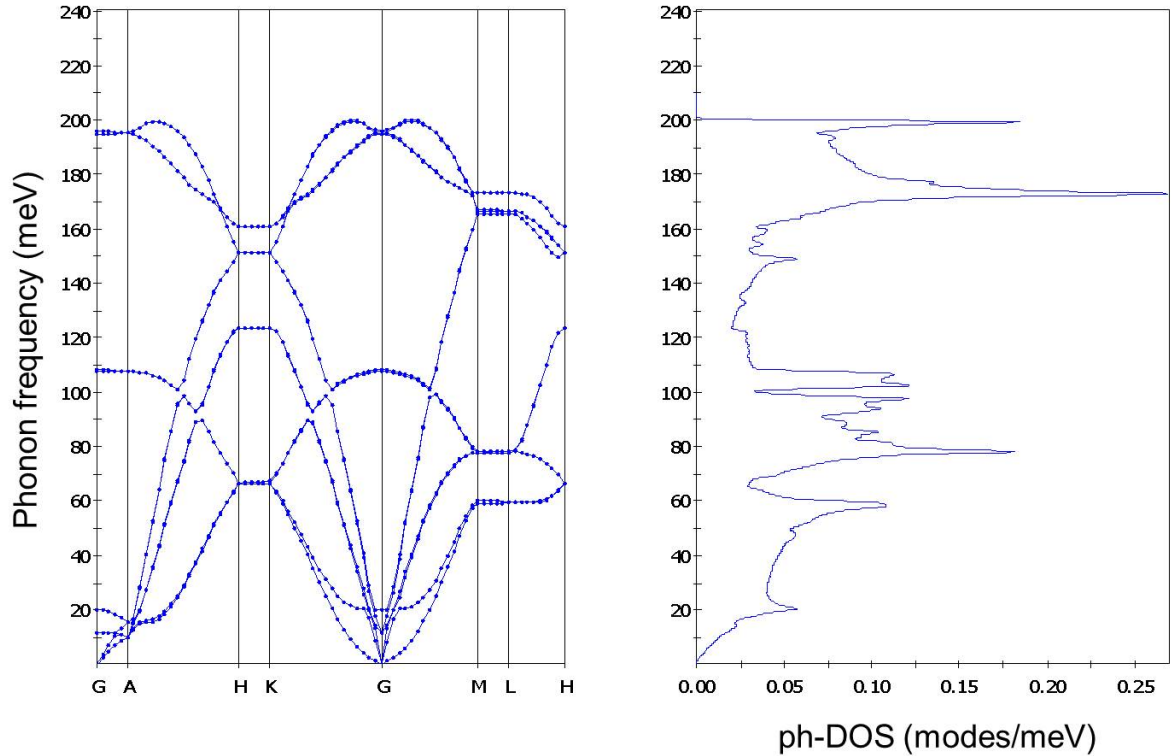

Fig. 1S

## B. Comparison of differential conductance (*dI/dV*) spectra and the phonon density of states (ph-DOS) for graphite

**Description:** Fig. 2S and 3S show the correlation between experimentally obtained *dI/dV* data at T = 4.2 K (bottom panel) and the theoretical values of phonon density of states (ph-DOS) for graphite (top panel) for two different devices, respectively. The green curves are the two-phonon density of states which describes the emission process of two phonons with the same energy but wave vectors of opposite directions. Vertical lines are eye guides to the positions of ph-DOS peaks.

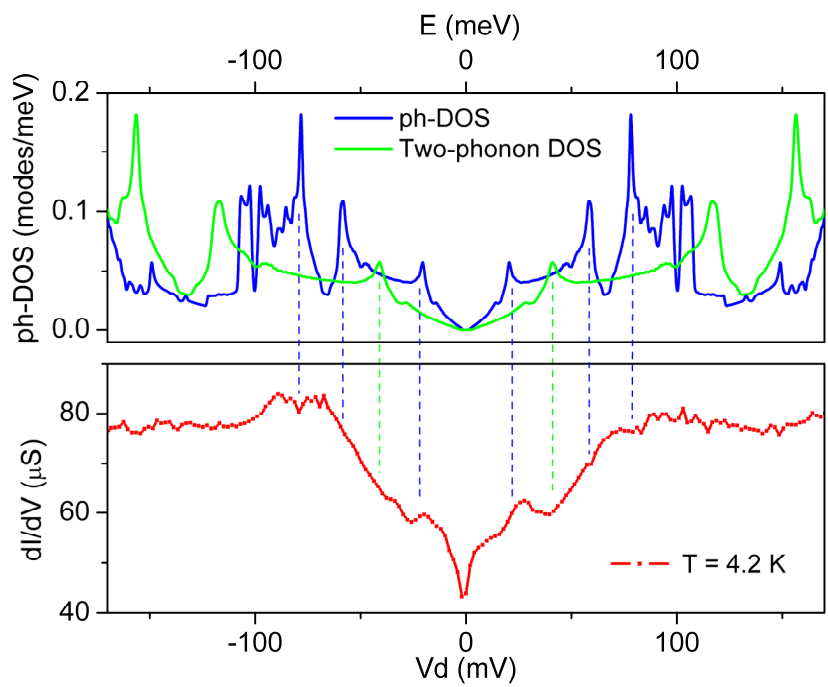

**Fig. 2S**

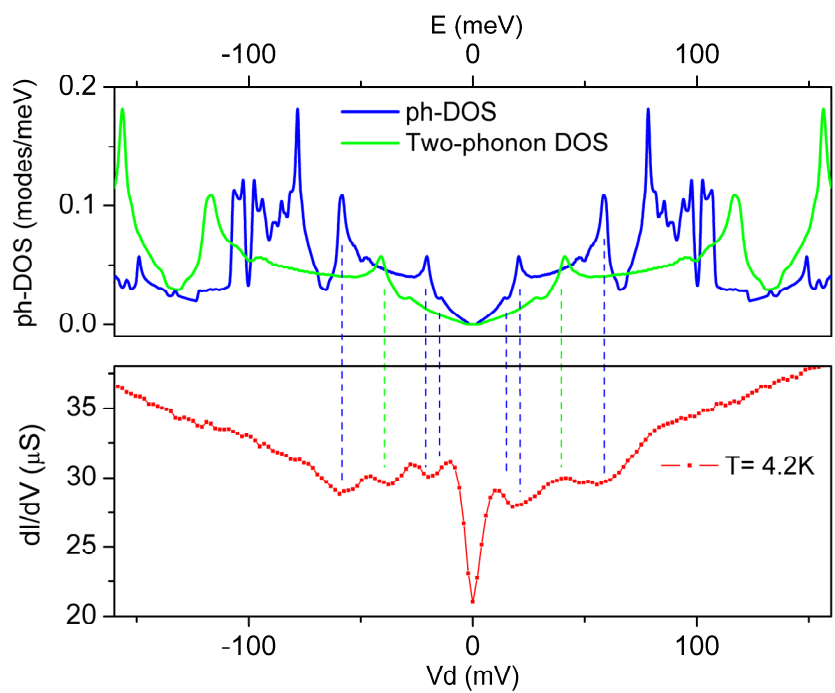

**Fig. 3S**